\documentclass[%
 reprint,
nofootinbib,
 amsmath,amssymb,
 aps,
]{revtex4-1}

\newcommand{\boldnabla}{\mbox{\boldmath$\nabla$}}

\usepackage{color}
\usepackage{graphicx}
\usepackage{dcolumn}
\usepackage{bm}
\usepackage[mathlines]{lineno}
\usepackage[perpage]{footmisc}
\begin{document}

\preprint{APS/123-QED}

\title{Fast chemical reaction in two-dimensional Navier-Stokes flow: \\ Initial regime}

\author{Farid Ait-Chaalal}
 \thanks{Corresponding author, new affiliation: \\ California Institute of Technology, California, \\ farid.ait-chaalal@gps.caltech.edu} 
\author{Michel S. Bourqui}
\author{Peter Bartello}
\affiliation{%
McGill University, Montr\'eal, QC, Canada
}%

\date{\today}

\begin{abstract}

This paper studies an infinitely fast bimolecular chemical reaction in a two-dimensional bi-periodic Navier-Stokes flow.
The reactants in stoichiometric quantities are initially segregated by infinite gradients. The focus is placed 
on the initial stage of the reaction characterized by a well-defined one dimensional material contact line between the reactants. Particular attention 
is given to the effect of the diffusion $\kappa$ of the reactants. This study is an idealized framework for 
isentropic mixing in the lower stratosphere and is motivated by the need to better understand the effect of resolution on 
stratospheric chemistry in Climate-Chemistry Models. 

Adopting a Lagrangian straining theory approach, we relate theoretically the ensemble mean of the length of the contact line, of the gradients along it and 
of the modulus of the time derivative of the space-average reactant concentrations (here called the chemical speed) to the joint probability density function
of the finite time Lyapunov exponent $\lambda$ with two times $\tau$ and $\widetilde{\tau}$. The time $\frac{1}{\lambda}$ measures 
the stretching time scale of a Lagrangian parcel on a chaotic orbit up to a finite time $t$, while $\tau$ measures it in the recent past before $t$, and 
$\widetilde{\tau}$ in the early part of the trajectory. We show that the chemical speed scales like $\kappa^{\frac{1}{2}}$ and that its time evolution
is determined by rare large events in the finite time Lyapunov exponent distribution. The case of smooth initial gradients is also discussed. The theoretical 
results are tested with an ensemble of direct numerical simulations (DNS) using a pseudospectral model.

\end{abstract}
                             
\maketitle

\section{\label{sec:level1}\ Introduction}

The stratospheric ozone chemistry resulting from Climate-Chemistry Models is thought to be sensitive to the spatial resolution. It was shown by
\cite{edouard1996} that the simulated spring ozone depletion inside the polar vortex is very sensitive to the horizontal grid size.  
However \cite{Searle1998a,Searle1998b} pointing out some flaws in the former work, suggested that resolution is not 
crucial for ozone depletion inside the polar vortex during cold enough winters because chlorine, the relevant catalyst for ozone destruction, 
is totally activated regardless the resolution. 
Nevertheless, they suggested that at the outer edge of the vortex (the surf zone), where 
mixing is important, the filamentary structures exhibited by chemical fields (e.g. \cite{Waugh1994}) are not represented by low resolution models. 
The deactivation of polar vortex chlorine by low-latitude nitrogen oxide, a process controlling
ozone concentrations at the outer edge of the mid-winter Arctic polar vortex, was studied numerically by \cite{Tan1998}. Assuming two dimensional 
mixing on isentropes on time scales smaller than two weeks and using reanalysis data to advect chemicals, they found that the production of chlorine 
was strongly dependent on the tracer diffusion coefficient. They proposed that the product's concentration scales like $\kappa^{p(t)}$ where 
$p(t)$ is a positive decreasing function of time which depends on initial conditions. This problem was addressed from a theoretical point of view by 
\cite{Wonhas02} which showed, for an infinitely fast bimolecular chemistry, that the function $p(t)$ is given by 
$1-D(t)/2$ where $D(t)$ is the box counting fractal dimension of the contact line between the reactants, defined as the zero isoline of 
a tracer $\phi$ equal to the difference between the two reactants' fields. Their main assumption on the geometric 
configuration of $\phi$ is that of an on/off field, which allows to link the slope of the tracers' variance spectrum to the box counting fractal 
dimension of the contact line, and the variance to the first moment of the modulus of $\phi$. 
This interesting approach is however limited by the lack of realism of the on/off fields assumption.

\paragraph*{}
Here, in the absence of this assumption, we propose to focus on the case where the contact line is a material line unaffected by diffusion 
(fractal dimension equals to one). This is true during the 
early stage of the reaction, before tracer filaments start to merge under the action of diffusion. To our knowledge, a detailed analysis of this regime
has not appeared in the literature despite its relevance to the atmosphere on time scales of several days to weeks.
We develop a mathematical framework which relates the effect of diffusion on the reactant concentration and its time evolution to the statistics
of Lagrangian straining properties (LSP) of advected parcels in the flow. This approach has been widely used
to describe the asymptotic decay of passive tracers in the Batchelor regime of turbulence or in chaotic advection 
(for Lagrangian straining theories and further developments, see \cite{Antonsen1996,Balkovsky99,Sukh02,Fereday02,Fereday04,Tsang05,Haynes05}). 
In addition, this approach was recently applied to the long term decay of fast reacting 
chemicals by \cite{Tsang09}. 

Our assumptions are those of a two dimensional statistically isotropic, homogeneous and stationary non linear Navier-Stokes flow. 
We use ensembles of direct numerical simulations, each one corresponding to a different initial condition on the vorticity, to verify the 
analytical relations between the LSP and chemistry. Although this flow gives a very simplified representation of stratospheric mixing, 
it can be argued that it is relevant for scales larger than approximately 40km (\cite{haynes97}). We vary the diffusion coefficient $\kappa$ of tracers 
to study the effect of resolution, employing a similar approach to \cite{Tan1998,Wonhas02}. This is justified by noting that the smallest scales of 
the flow are determined by the balance of advective and  diffusive processes and thus scale like $\kappa^{\frac{1}{2}}$.
Considering that small-scale tracer structures are generated by the large-scale field, the viscosity of the field is chosen larger
than the diffusion $\kappa$. Hence, tracers evolve in a smooth velocity field, which allows to differentiate it at the 
tracers' small scales and interpret their behavior in the framework of Lagrangian chaos.  It has been shown in \cite{ngan1,ngan2} 
that the concept of chaotic advection, where a spatially coarse flow produces chaotic tracer trajectories, was applicable to two-dimensional 
mixing in the stratospheric surf zone. In addition it has been argued (\cite{bartello00}) that in barotropic, beta-plan two-dimensional turbulence, relatively 
coarse velocity fields reproduce quite accurately the fine structures of the tracer field when the spectrum of energy is steeper than $k^{-3}$, which is 
relevant both in the stratosphere (\cite{koshyk01}) and in the enstrophy cascade in two-dimensional turbulence (\cite{kraichnan1967}). 

\begin{figure*}
  \centering
  \includegraphics[scale=0.6]{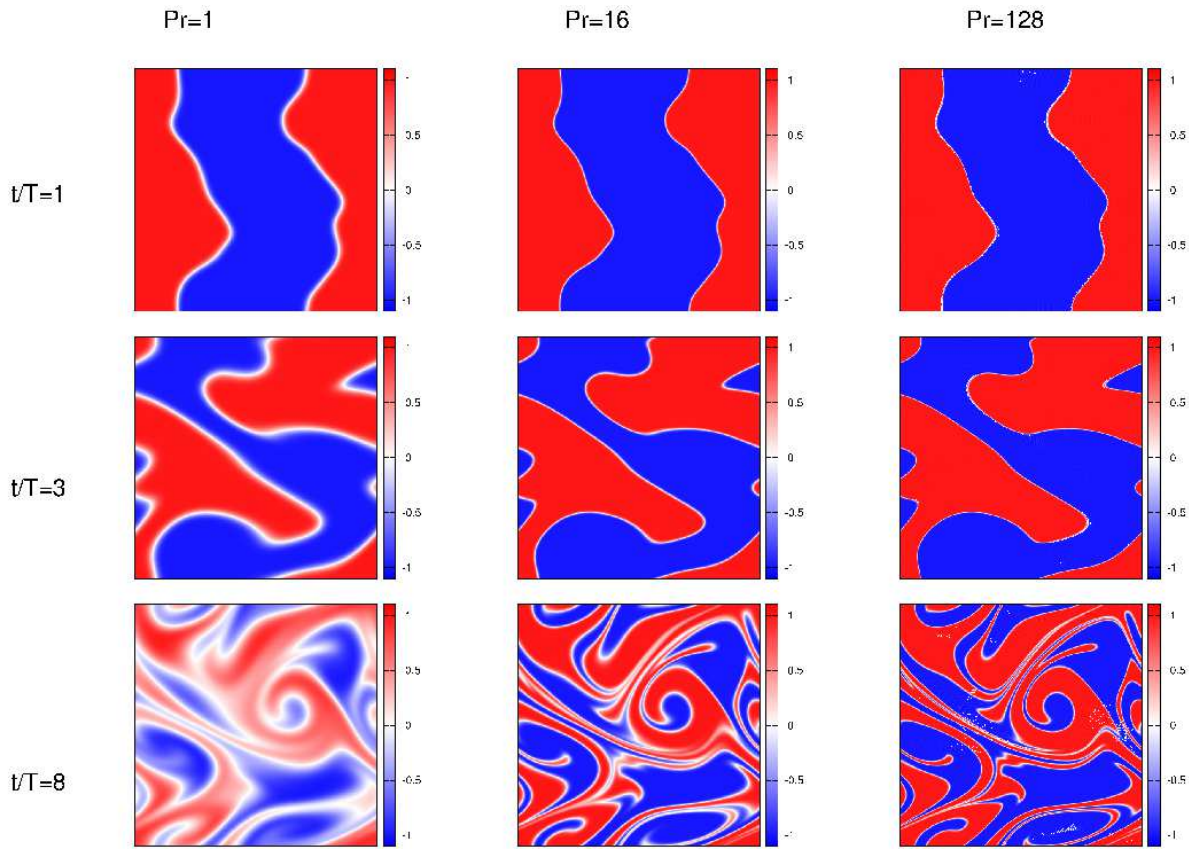}
  \caption{(Color online)  \small Reactant fields in a bi-periodic domain $[-\pi,\pi]^2$. Red (or light grey), positive values, and blue (or dark grey), negative values, 
  show the two reactants A and B. From left to 
  right $\Pr=1,16,128$ and from top to bottom $\frac{t}{T}=1,3,8$. The Prandtl number $Pr$ is defined as the ratio of the viscosity of the fluid to the 
  tracer diffusion. Since the viscosity is fixed, an increasing $Pr$ means a decreasing diffusion. $T$ is the integral time scale of the flow.} 
\label{F:show}
\end{figure*}

We focus on the initial regime of an
 infinitely fast chemical reaction between two segregated reactants in stoichiometric quantities. The main emphasis is placed on 
the case where the reactants are initially separated by a sharp gradient, while the case of a smooth gradient is briefly discussed. 
Figure $\ref{F:show}$ illustrates 
this regime. With $T$ the integral time scale of the flow, the contact line does not depend on diffusion at 
$\frac{t}{T}=1$ and  $\frac{t}{T}=3$, but gradients become clearly smoother when diffusion increases. The time span of this regime depends on the 
diffusion: at $\frac{t}{T}=8$ the contact line seems to be the same for Prandtl numbers $Pr=16$ and $Pr=128$ but is clearly different for $Pr=1$. The Prandtl number is 
defined as the ratio of the tracer diffusion $\kappa$ with the fluid viscosity $\nu$ and when the diffusion
is larger, filaments merge earlier, making the contact line dependent on diffusion at a smaller time. 

This approach is relevant to the chlorine deactivation at the outer edge of the winter time polar vortex, which is very fast compared to advective and
diffusive time scales (\cite{Tan1998}). It is also of general interest in isolating and investigating the effect of two-dimensional turbulent mixing on chemical reactions. 
A separate paper, in preparation, will focus on the case of 
a more complicated contact line (box counting fractal dimension between 1 and 2), which corresponds to the intermediate and time asymptotic regime. 

\paragraph*{}
This paper is organized as follows. Section II. describes our approach and methodology. We show that with infinitely fast chemistry, 
average concentrations of reactants and product are simple linear functions of the first moment of the modulus of the passive tracer concentration 
$\phi$ defined as the difference between the reactant fields. This approach is rather general in the study of infinitely fast bimolecular reactions 
(\cite{Corrsin1957,Sokolov1991,Wonhas02,Tsang09}). In particular, this implies that the reaction
is controlled by the diffusive flux across the isoline $\phi=0$, noted $\mathcal{L}$. The importance of the behavior
of $\mathcal{L}$ for chemistry in complex flows or complex geometric configurations of chemical fields has been highlighted in 
\cite{Giona01,Martinand07} respectively. This section also describes the numerical model and the simulated flow, including the 
spatial configuration and the probability density function (pdf) of the FTLE. Section III. describes the theoretical and numerical results. 
We derive analytical expressions for the lengthening of $\mathcal{L}$, for the gradients advected along $\mathcal{L}$ and finally for the diffusive 
flux across $\mathcal{L}$, the latter being equal to the time derivative of the space-average reactants concentrations. We compare the theory to ensembles of numerical
simulations. Finally, conclusions are drawn in section IV.

\section{Methodology}

\subsection{The limit of infinite chemistry}

We consider the bimolecular chemical reaction $A + B \longrightarrow C$ with $C_A$, $C_B$ and $C_C$, 
the concentrations of A, B and C respectively. Eulerian equations describing the evolution of 
$C_{i}(\mathbf{x},t)$, $i=A,B,C$, in the flow $\mathbf{u}=(u,v)$ are:
\begin{subequations} \label{E:gp}
  \begin{gather}
    \frac{\partial{C_A}}{\partial {t}} + \mathbf{u}\cdot \nabla {C_A} = 
      \kappa \nabla^2  {C_A} -k_c {C_A}{C_B} \label{E:gp1} \\
    \frac{\partial{C_B}}{\partial {t}} + \mathbf{u}\cdot \nabla {C_B} = 
      \kappa \nabla^2  {C_B} -k_c {C_A}{C_B} \label{E:gp2} \\
    \frac{\partial{C_C}}{\partial {t}}  + \mathbf{u}\cdot \nabla {C_C} = 
      \kappa \nabla^2  {C_C} +k_c {C_A}{C_B} \label{E:gp3} \mbox{ ,}
  \end{gather}
\end{subequations}
where $k_c$ is the chemical reaction rate and $\kappa$ the diffusion, which is assumed equal for all tracers.
The quantity $\phi=C_A-C_B$ is a passive tracer which obeys the simple advection-diffusion equation 
\begin{equation}\label{E:tracer}
  \frac{\partial{\phi}}{\partial {t}} + \mathbf{u}\cdot \nabla {\phi} = \kappa \nabla^2  {\phi}                                                    
\end{equation}
We assume that $\phi$ has zero spatial average, which is equivalent to having the reactants in
stoichiometric balanced ratio. 
Under the fast chemistry hypothesis ($k_c\longrightarrow \infty$), we can assume without loss of generality that the reactants 
$A$ and $B$ are segregated (i.e. A and B do not overlap spatially). In fact, even if they are co-located at time $t=0$, 
they can not coexist at a later time $t > 0$ since they  
react instantaneously where both fields are together non-zero. 
It follows that:
\begin{equation}
  \left\{\begin{matrix}
    C_A(\mathbf{x},t)= \phi(\mathbf{x},t) & \mbox{ and } C_B(\mathbf{x},t)=0 & \mbox{ if } \phi(\mathbf{x},t) > 0 \\ 
    C_B(\mathbf{x},t)=-\phi(\mathbf{x},t) & \mbox{ and } C_A(\mathbf{x},t)=0 & \mbox{ if } \phi(\mathbf{x},t) < 0 
  \end{matrix}\right.
\end{equation}
Defining with an over-bar the average over the whole domain, we have:
\begin{subequations} \label{E:tot}
  \begin{gather}
    \overline{C_A} = \overline{C_B}=\frac{\overline{\left |\phi\right |}}{2} \label{E:tot1} \\
    \overline{C_C} = \frac{\overline{\left | \phi(t=0)\right |} - \overline{\left |\phi\right |} }{2}  \mbox{.}
  \end{gather}
\end{subequations}

Consequently, studying the decay of the reactants of an infinitely fast chemical reaction 
in stoichiometric balanced ratio is equivalent to studying the decay of the first moment 
of the modulus of a passive tracer $\phi$ of zero spatial average. 
For an incompressible flow, it can be shown with the divergence theorem that the decay rate of the total reactant quantity for an infinite reaction 
equals the diffusive flux across the contact line between $A$ and $B$, namely $\mathcal{L}=\{\mathbf{x}\lvert\phi(\mathbf{x})=0\}$, oriented in a counterclockwise
direction around the area where $A$ is located:
\begin{eqnarray}\label{E:difflux}
  \mathcal{A} \frac{\overline{dC_A}}{dt} = \mathcal{A} \frac{\overline{dC_B}}{dt}=\frac{1}{2} \mathcal{A} \frac{d \overline{\left | \phi \right |}}{dt}&=&
  - \kappa \int_{\mathcal{L}(t)} \nabla \phi \cdot \mathbf{n}\,dl \nonumber \\ &=& - \kappa \int_{\mathcal{L}(t)} \left|\nabla \phi \right| \,dl \mbox{,}
\end{eqnarray}
where $\mathcal{A}$ is the total area of the domain and $ \mathbf{n}$ the vector normal to $\mathcal{L}$ and pointing outside the area where $A$ is located.
The contact line is by definition an isoline of $\phi$, which gives the last equality in ($\ref{E:difflux}$).
Hereafter, $-\frac{d \overline{\left | \phi \right |}}{dt}$ is called the chemical speed.

\subsection{The numerical model}

\subsubsection{The flow}

The numerical model integrates the vorticity equation:
\begin{equation}\label{E:vorticity}
  \frac{\partial{\omega}}{\partial{t}}+\mathbf{u} \cdot \boldnabla \omega=F-R_0 \omega +  \nu \nabla^2 \omega
\end{equation}
where $\omega=\bm{\nabla} \times \bm{u}$ is the vorticity, $F$ the forcing term, $R_0$ the Rayleigh friction and $\nu$ the viscosity. 
The equation is integrated in a bi-periodic domain $(x,y) \in   [-\pi,\pi]^2$ on a  $512 \times 512$ grid
using the pseudo-spectral method. 
The fast Fourier transforms are provided by FFTW (\cite{fftw}). The Fourier series are truncated at 
$K_{max} = 512/3$ to avoid aliasing. The time stepping algorithms are leap-frog for the advection and Crank-Nicholson for 
the viscosity. The computational mode is dissipated by a weak Robert filter with parameter $0.001$. 
The forcing term $F$ has the following form in Fourier space:

\begin{equation}
  F_\mathbf{k}=
  \left\{\begin{array}{lcc}
    0.002 \mbox{ if } \mathbf{k}=(\pm 3,0) \mbox{ and } \mathbf{k}=(0,\pm 3)\\ 
    0 \mbox{     otherwise }
  \end{array}\right.
\end{equation}
The energy tends to concentrate in the largest scales of the flow because of the inverse energy cascade 
inherent to two-dimensional turbulence. As a consequence, we use a Rayleigh friction term with $R_0=0.0002$
in the vorticity equation ($\ref{E:vorticity}$) to balance the injection of energy through $F$. The viscosity is $\nu \simeq 5.57 \times 10^{-4}$ 
and results in a Reynolds number $Re$ of the order of $10^3$. It has deliberately been chosen to be relatively low for reasons explained in 
the Introduction.

A snapshot of the vorticity field is depicted in figure $\ref{F:mapFTLE}$ (top left). 
With brackets for an ensemble average, the flow has an RMS velocity $\langle\overline{\mathbf{u} \cdot \mathbf{u}}\rangle^\frac{1}{2} \simeq 0.08$ and a mean enstrophy 
$Z=\frac{1}{2}\langle\overline{\omega^2}\rangle \simeq 0.009$, which corresponds to an advective time scale $T = Z^{-\frac{1}{2}} \sim 10$ (\cite{bartello00}).
Hereafter, $T$ is used to normalize time and can also be estimated from the mean strain rate ${\langle \overline{S} \rangle}$, where 
$S = \left(\left(\frac{\partial u}{\partial x}\right)^2+\frac{1}{4}\left(\frac{\partial u}{\partial y} + \frac{\partial v}{\partial x} \right)^2\right)^\frac{1}{2}$
(here expressed in Cartesian coordinates for an incompressible flow). In two dimensional turbulence, we have $\langle \overline{(2S)^2} \rangle=\langle \overline{\omega^2} \rangle$
(e.g. \cite{lapeyre02}). The distribution of the strain is close to a Rayleigh distribution 
\footnote[1]{It would be exactly a Rayleigh distribution if the velocity derivatives had Gaussian statistics and where statistically independent.}, 
as a consequence we have $\langle \overline{S^2} \rangle \approx \frac{4}{\pi} \langle \overline{S} \rangle^2$.  Finally, we have $T \approx \sqrt{\frac{\pi}{2}}\frac{1}{2\langle \overline{S} \rangle}$. 
The term $ \sqrt{\frac{\pi}{2}}$ being of the order of unity, the turnover time can be evaluated from $\frac{1}{2 \langle \overline{S} \rangle}$. The mean strain rate is about $0.05$ 
in our flow (see figure $\ref{F:plotpdf}$ for the whole distribution), which gives approximately the same estimate as $Z^{-\frac{1}{2}}$ for $T$.

\subsubsection{Finite Time Lyapunov Exponents (FTLE)}

\paragraph{Definition and properties}

The finite time Lyapunov exponent is defined as the rate of exponential increase of the distance 
between the trajectories of two fluid parcels that are initially infinitely close. If $\bm{\delta l}(t)$ is the distance at time $t$ between two 
parcels that start at $\bm{x}$ and  $\bm{x+\delta l_0}$ at time $t=0$, then the FTLE $\lambda(\bm{x},t)$ at $\bm{x}$ over the time interval $t$ is
\begin{equation}\label{E:ftle_def}
 \lambda(\bm{x},t)=\frac{1}{t} \max_{\alpha} \underset{\bm{|\delta l_0|} \to 0}{\lim} \big\{ \ln{\frac{\bm{|\delta l(t)|}}{\bm{|\delta l_0|}}} \big\} \mbox{   ,}
\end{equation}
where the maximum is calculated over all the possible orientations $\alpha$ of $\bm{\delta l}_0$. The unit vector with the 
orientation $\psi_{+}(\mathbf{x},t)$ of $\bm{\delta l}_0$ at the maximum is called a singular vector and we note it
$\bm{\psi_+}(\mathbf{x},t) \equiv (\cos \psi_{+},\sin \psi_{+})$. It defines a Lagrangian straining direction. 
It follows from ($\ref{E:ftle_def}$) that the FTLE converges to the strain rate as $t \rightarrow 0$. 
For large times, the large deviation theory
suggests that the FTLE pdf $P_{\lambda}$ in chaotic flows without KAM (Kolmogorov, Arnold, and Moser) surfaces
(\cite{ottino02}) can be well approximated by:
\begin{equation}\label{E:ftle_pdf}
  \widetilde{P}_{\lambda}(t,\lambda)=\sqrt{\frac{t G''(\lambda_0)}{2\pi}} \exp(-tG(\lambda))\mbox{,}
\end{equation}
where $G(\lambda)$, the Cramer or rate function, is concave with its minimum at $\lambda_0$ satisfying $G(\lambda_0)=G'(\lambda_0)=0$. 
Moreover, $\lambda_0$ is the infinite-time Lyapunov exponent: 
$\underset{t \to \infty}{\lim} P_{\lambda}(t,\lambda)=\delta(\lambda_0-\lambda)$ where $\delta$ is the Dirac delta function. The convergence
of the Lypunov exponents is very slow and typically algebraic in time (\cite{Tang1996}). The form ($\ref{E:ftle_pdf}$) 
has been numerically verified and is widely used to approximate the asymptotic form of FTLE pdfs in simple ergodic flows with
chaotic advection (e.g \cite{Antonsen1996,Haynes05,Tsang09}).

\paragraph{Computation and description}

The distance $\bm{\delta l}$ between two trajectories initially infinitely close is solution of 
\begin{equation}\label{E:evol_dl}
\frac{d\bm{\delta l}}{dt}-\mathbf{S}.\bm{\delta l}=0 \mbox{,}
\end{equation}
where the tensor $\mathbf{S} = \nabla \mathbf{u}(\mathbf{X},t)$ is the velocity gradient tensor
along a trajectory $\mathbf{X}(\mathbf{x},t)$. The distance $\bm{\delta l}$ can be calculated by $\bm{\delta l}=\mathbf{M}\bm{\delta l}_0$,
where the resolvent matrix $\mathbf{M}$ is solution of $\frac{d\mathbf{M}}{dt}-\mathbf{S}\mathbf{M}=0$ and is equal to the identity at $t=0$. 
The finite time Lyapunov exponent $\lambda$ is $\frac{1}{2 t}$ times the $\log$ of the largest eigenvalue of $[\mathbf{M}^{T}\mathbf{M}]$, 
with the singular vector $\bm{\psi}_{+}$ the associated eigenvector. The FTLE are obtained  using the method described in \cite{abr02} from the trajectories 
computed offline using a fourth order Runge-Kutta scheme with a trilinear interpolation on the velocity field. The time step is $0.1$, which corresponds 
to a hundredth of the turnover time. The tensor $\mathbf{S}$ is calculated along the trajectories to obtain  $\mathbf{M}$ and consequently $\lambda$ 
and $\bm{\psi}_{+}$.

\begin{figure}
\centering
\includegraphics[scale=1.1]{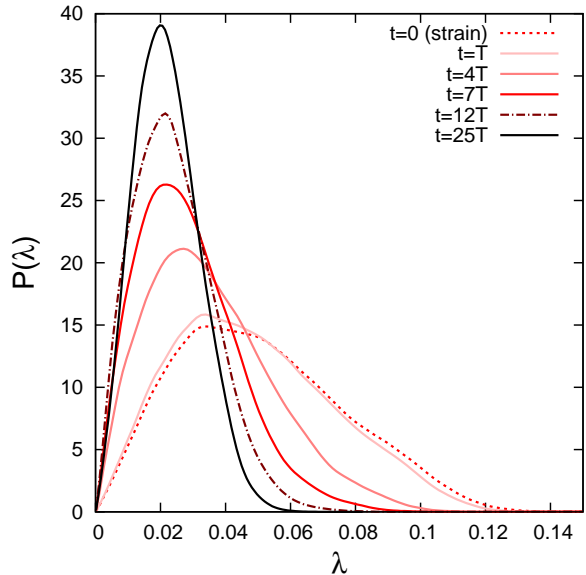}
\caption{(Color online)  \small Density $P_{\lambda}$ of the finite-time Lyapunov exponents shown at different times between $t=0$ and $t=25T$. As the time increases, the density
                shifts toward smaller values.}
\label{F:plotpdf}
\end{figure}

\begin{figure}
\centering
\includegraphics[scale=1.1]{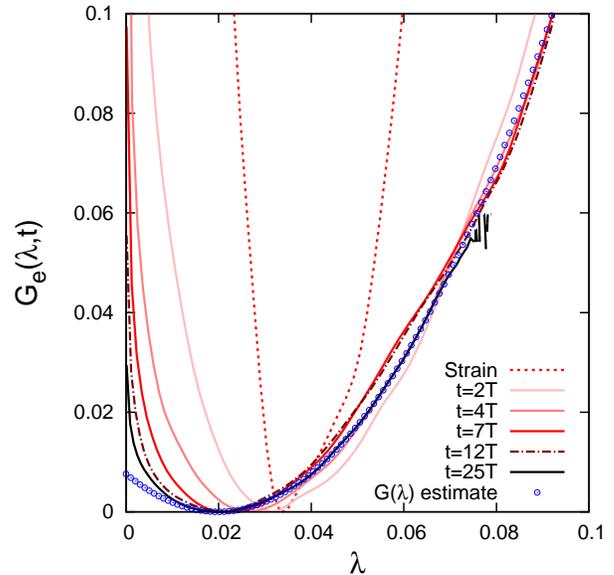}
\caption{(Color online)   \small  Function $G_e(\lambda,t)$ plotted at different times ($0 < t/T < 25$).  $G_e$ is defined such that $P_{\lambda}$ - 
plotted in figure $\ref{F:plotpdf}$ - can be written $\propto {-t G_e(\lambda,t)}$ with $\underset{\lambda}{\min} G_e=0$. 
As the time increases, the function $G_e$ shifts toward smaller values. We note the asymmetry of $G_e$ 
and the faster convergence for FTLE larger than their ensemble mean. The time asymptotic form of $G_e$ is the Cramer function $G$ corresponding to the
longtime FTLE pdf $P_{\lambda}$. An estimate of $G$ is given by the blue circles, using a method detailed in the text.}
\label{F:cramerplot}
\end{figure}

We estimate the FTLE pdfs as normalized histograms over 100 realizations of the flow, differing by their initial vorticity field. We initialise a
trajectory at every grid point of our $512^2$ domain, which results of a total of about $26 \times 10^6$ trajectories calculated. Each realization is run for a time 
span of $25 T$. The FTLE pdfs are shown
at different times in figure $\ref{F:plotpdf}$. The variance of the FTLE decreases with time while the peak of the distribution converges 
toward $\lambda_{max}\sim 0.02$. The FTLE are significantly smaller than the strain rates probably because of vorticity that inhibits the
stretching of material elements (\cite{McWilliams1984}) and because of the reorientation of the local strain axis along a trajectory 
(e.g. \cite{Chertkov1995}).
In order to estimate whether these pdfs are asymptotically well approximated by ($\ref{E:ftle_pdf}$), we define
\begin{equation}\label{E:evol_cramer}
	G_e(\lambda,t)=-\frac{\ln(P_{\lambda}(t,\lambda))}{t}+\frac{\ln t}{2t}+\frac{A_e(t)}{t} \mbox{, }
\end{equation} 
where $A_e(t)$ is chosen such that $\underset{\lambda}{\min} G_e(\lambda,t)=0$.
Figure $\ref{F:cramerplot}$ shows the time evolution of $G_e$. The convergence for large values of $\lambda$, typically larger
than the mean $\langle \overline{\lambda} \rangle$ is satisfactory. However, the convergence for small values is much slower. It is 
particularly difficult to get the Cramer function for small values of $\lambda$ (\cite{Vanneste10}). This is not a concern for the 
present study because only values of $\lambda$ larger than their average are relevant. Nevertheless,  
we can get an estimate of the Cramer function assuming it is symmetric, as represented on figure $\ref{F:cramerplot}$. We have fitted the average of the Cramer
function for times larger that $15T$ for values of $\lambda$ larger than $0.02$ with a second order polynomial (gaussian approximation) and obtained
the estimate for values smaller than $\langle \overline{\lambda} \rangle$ using symmetry.   

\begin{figure*}
\centering
\includegraphics[scale=1.2]{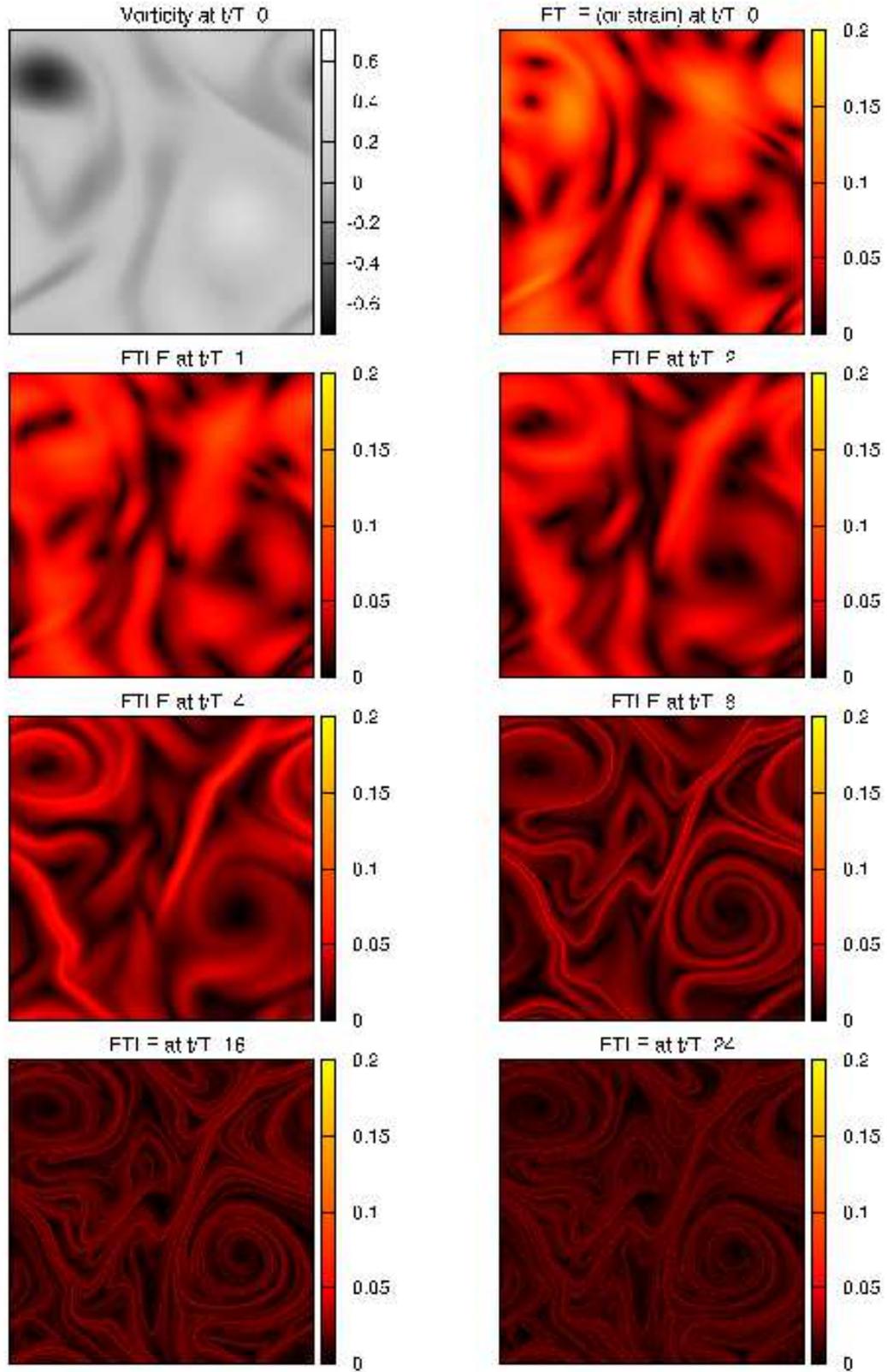}
\caption{(Color online)  \small Maps of FTLE calculated at different times and displayed at starting locations of trajectories in the bi-periodic domain $[-\pi,\pi]^2$. 
In the top row are shown the vorticity (left) and the strain (right) at $t=0$. In the following panels, ordered from left 
to right and top to bottom are shown the FTLE maps for $\frac{t}{T}=1,2,4,8,16,24$.}
\label{F:mapFTLE}
\end{figure*}

The FTLE maps are shown in figure $\ref{F:mapFTLE}$. For small times the strain field is dominated by large scales because two dimensional dynamics do not allow
a cascade of energy toward smaller scales.
However filamentary structures appear shortly, becoming finer and finer until they reach the resolution of the Eulerian model (the trajectories are initiated at
every grid point of the Eulerian model). It is interesting to note that 
we get very similar structures as \cite{lapeyre02}, despite our much coarser velocity field. In particular, in the direction perpendicular to the FTLE filaments, 
it can be observed than initially close trajectories can have totally different FTLE. This might be a manifestation of chaotic advection.

It has been argued, in ergodic systems, that the singular vectors converge exponentially in time (\cite{Gol87}), faster than the Lyapunov exponents, whose 
convergence is algebraic (\cite{Gol87,Tang1996,lapeyre02}). The ``freezing'' of the large scale patterns in the FTLE maps (figure $\ref{F:mapFTLE}$) may be
interpreted as a manifestation of the convergence of the singular vectors. In fact, \cite{Tang1996} argued that in ergodic and conservative chaotic dynamical 
systems, the Lyapunov exponents varies slowly along lines (the $\mathbf{\hat{s}}$ lines) which defines the stable direction in which neighboring points 
asymptotically converge. The filamentary structures in figure $\ref{F:mapFTLE}$ may be interpreted as being these $\mathbf{\hat{s}}$ lines. This has
been verified experimentally through the computation of the singular vectors (not shown), their convergence being particularly fast for trajectories
originating in areas of the flow dominated by strain. 

In the theoretical developments of part III.B and III.C, we will neglect the time evolution of the Lyapunov vectors and will only take into account the
time evolution of the Lyapunov exponents. 

\subsubsection{The tracers}

The passive scalar $\phi$ is integrated with equation ($\ref{E:tracer}$), using the same numerical scheme as for the vorticity.  
The numerical simulations are performed for eight different Prandtl numbers $Pr= \frac{\kappa}{\nu}=2^i \mbox{ for } 0 \leq i \leq 7$.
Consequently the Peclet number $Pe = Pr Re$, which measures the ratio of the advective to the diffusive time scale, ranges from
$10^3$ to $10^5$.

We use two different initial conditions on the tracer for $(x,y)\in[-\pi,\pi]^2$:
\begin{align}
    \phi(x,y,t=0)&=A_0 \operatorname{sgn} x & \notag \\& \mbox{ for infinite initial gradients  } & \label{E:init-trac_1} \\   
    \phi(x,y,t=0)&=A_0 \frac{\pi^2}{4}\cos x \cos y & \notag \\& \mbox{ for smooth initial gradients} & \label{E:init-trac_2}  \mbox{,}  
\end{align}
where $\operatorname{sgn} x$ is the sign of $x$ and $A_0$ is twice the initial domain average concentration of both $A$ and $B$ in the box.
The first initial condition allows represents the case of sharp (actually infinite) gradients separating
areas of well mixed reactants and the second one the case of smooth gradients.\\

\subsection{The ensemble analysis}

For each value of Prandtl number and for each initial condition, we run an ensemble of 34 
simulations (or members). Each member is defined by different initial condition on the vorticity, taken from a long simulation of the statistically stationary 
flow solution of ($\ref{E:vorticity}$).

\section{Theoretical and numerical results}

Our goal here is to describe and understand the initial evolution of the first moment of $|\phi|$. In other words, we would like to integrate 
($\ref{E:difflux}$). We first consider how a material line stretches in a Lagrangian framework (III.A), and then how gradients on the contact line
evolve under the action of both the diffusion and the flow along a Lagrangian trajectory (III.B). Paragraph III.C deals with the 
chemical speed. We focus on the initial condition where the reactants are separated by a sharp gradient before discussing the case of smoother 
gradients (III.D).

\subsection{Lengthening of the contact line $\mathcal{L}$}

\subsubsection{Theory}

We consider a line element  $\mathbf{\boldsymbol \delta l_{0}}$ along the contact line 
$\mathcal{L}(t=0) \equiv \mathcal{L}_0$. Its coordinates are $\delta l_{0} (\cos \alpha,\sin \alpha)$. The angle $\alpha$ is the initial orientation of the line element. 
It is transformed at time $t$ into an element $\bm{\delta l}=\mathbf{M} \bm{\delta l_{0}}$ whose norm is:
\begin{equation}\label{E:coorddl}
\begin{array}{ll}
  \left| \bm{\delta l} \right| &= \left[\bm{\delta l_{0}}^{T}\mathbf{M}^{T}\mathbf{M}\bm{\delta l_{0}}\right]^{\frac{1}{2}}\\ &=\left | \bm{\delta l_{0}} \right | 
  \left[e^{2 \lambda t}\cos^2(\psi_{+}-\alpha) + e^{-2 \lambda t}\sin^2(\psi_{+}-\alpha) \right ]^{\frac{1}{2}} \mbox{.}
\end{array}
\end{equation}
The resolvent matrix $\mathbf{M}$ was introduced in II.B.2.b. The angle  $\alpha$ and consequently the angle $\gamma \equiv \psi_{+}-\alpha$ between 
the initial orientation and the singular vector can be assumed uniformly distributed between $0$ and $2\pi$ and statistically independent of the chaotic orbit because the contact 
line is chosen arbitrarily with respect to the flow.
Integrating over the Lyapunov exponent $\lambda$,  the angle $\gamma$ and the 
initial contact line, gives the 
ensemble average $\langle L \rangle$ of the length $L$ of $\mathcal{L}$ (brackets are for ensemble averages). With $P_{\lambda}$
the probability density distribution of $\lambda$, we have:
\begin{align}
\langle L \rangle & =L_{0}\int_{\lambda=0}^{\infty} \int_{\gamma=0}^{2\pi} &
\left[e^{2 \lambda t}\cos^2 \gamma + e^{-2 \lambda t}\sin^2\gamma\right]^{\frac{1}{2}} \notag \\
&& P_{\lambda}(t,\lambda)\, d\lambda\, \frac{d\gamma}{2\pi} \mbox{.} \label{E:length_contact} 
\end{align}
The length $L_0$ is the initial length of the contact line. 
Equation ($\ref{E:length_contact}$) gives the actual length with no diffusion. Given the chaotic and closed (periodic) nature of the flow,
we can only neglect diffusion as long as the contact line has not folded on itself. Indeed, when two filaments of $\mathcal{L}$ are brought 
together at a distance smaller than the diffusive cutoff, they merge under the action of diffusion. 
The time span of the regime where ($\ref{E:length_contact}$) is expected to be valid can be approximated with the mix-down time $T_{mix}$ from the the 
largest scale $L$ of the flow to the diffusive cutoff $L_{\kappa}$ which is, 
according to \cite{th97}, $\frac{1}{\lambda} \ln(L/L_{\kappa})$, where $\lambda$ is the thinning rate of a fluid element, i.e the Lyapunov exponent.
It follows that $T_{mix}$ depends on the trajectory we are considering. To obtain an estimate of $T_{mix}$, we use $\lambda \approx \langle \overline{S} \rangle $ and 
$L_{\kappa} \approx \sqrt{\frac{\kappa}{\langle \overline{S} \rangle}}$:
\begin{equation}\label{E:tau_mix}
	T_{mix} \approx T\ln Pe = T\ln Re Pr \mbox{,} 
\end{equation}

The length $\langle L \rangle$ can be approximated by $L_{E}$ when we neglect the sine term 
in ($\ref{E:length_contact}$), i.e. when the contact line elements have equilibrated with the flow: their length converge to a function that
grows exponentially at a rate given by the FTLE, the initial orientation $\alpha$ of the contact line being ``forgotten''.  
This is valid for $t\gg\frac{1}{4 \langle \overline{S} \rangle} \approx \frac{T}{2}$.
\begin{align}\label{E:length_contact_s}
\langle L \rangle \underset{t \gg \frac{T}{2}}{\sim} L_{E} &=L_{0} \int_{\lambda=0}^{\infty} \int_{\gamma=0}^{2 \pi} 
					    P_{\lambda}(t,\lambda) \left | \cos \gamma \right | e^{\lambda t}d\lambda\,\frac{d\gamma}{\pi} \notag \\ 
 &= \frac{ 2 L_0}{\pi} \int_{0}^{\infty}    P_{\lambda}(t,\lambda) e^{\lambda t} d\lambda \mbox{.}
\end{align}
If we assume $P_\lambda(t,\lambda) \propto e^{-G_e(\lambda,t) t}$, with $G_e$ a concave positive function, integrating ($\ref{E:length_contact_s}$) with
the steepest descent method, we obtain:
\begin{equation}\label{E:length_contact_ss}
L_{E} \propto  \int_{0}^{\infty} e^{[\lambda  - G_e(\lambda,t)]t} d\lambda = e^{\max_{\lambda} [\lambda-G_e(\lambda,t)] t} \mbox{.}
\end{equation}
Asymptotically, we have:
\begin{equation}
  \label{E:long_inter}
  L_E \asymp e^{\lambda_1 t} \mbox{,}
\end{equation}
where
\begin{equation}
  \label{E:legendre1}
  \lambda_1 = \max_{\lambda} [\lambda-G(\lambda)]
\end{equation}
is the Legendre transform of $G$ evaluated with argument one. The value of $\lambda_1$ from our numerical estimate of $G$ (figure $\ref{F:cramerplot}$) is
$0.027$.

\subsubsection{Numerical results}

\begin{figure} 
\centering
\includegraphics[scale=1.1]{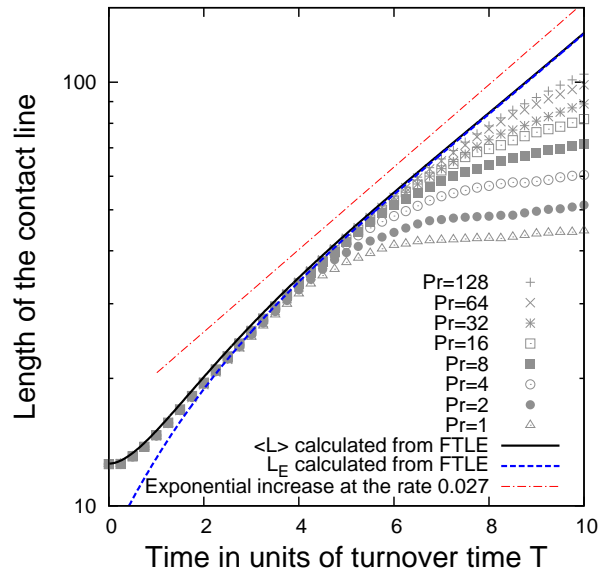}
\caption{(Color online)  \small Ensemble average of the length of the contact line (infinite initial gradient case). The symbols correspond to ensemble averages 
of DNS for different Prandtl numbers $Pr=1,2,4,8,16,32,64$ and $128$. The black solid line corresponds to $\langle L \rangle$, 
as estimated from the FTLE pdf using ($\ref{E:length_contact}$), and the bold blue dashed line to $L_{E}$ as estimated using the FTLE pdf with simplified 
expression ($\ref{E:length_contact_s}$). The light red dot-dashed line corresponds 
to an exponential increase at a rate $\lambda_1 = \underset{\lambda}{\max} [\lambda-G(\lambda)] \approx 0.027$, which corresponds to the asymptotic behavior in 
the inviscid limit ($\ref{E:long_inter}$). The latter has been shifted vertically for clarity. We note the log scale in the y-axis.}
\label{F:plotlength}
\end{figure}

The theoretical predictions $\langle L \rangle$ and $L_{E}$ are compared to the numerical calculations in figure $\ref{F:plotlength}$.
The integration of ($\ref{E:length_contact}$) using our numerical estimate of $P_{\lambda}$ reproduces very 
accurately the initial lengthening of the contact line for $t \lesssim T_{mix}(Pr)$. The derivative of the mean length of the contact 
line at $t=0$ is $0$ because contracting line elements statistically compensate with stretching line elements due to randomness of $\alpha$.
The inflection of $\ln\langle L \rangle$ around $\frac{t}{T}=2.5$ is due to two opposite effects: the equilibration of the contact line with the flow 
accelerates the growth of the line, while the shift of the FTLE pdf toward smaller values decelerates it, as shown by the $L_E$ curve. 

As seen on figure $\ref{F:plotlength}$, $\langle L \rangle$ and $L_{E}$ have a behavior very close to an exponential increase at the rate 
$\lambda_1 \approx 0.027$ after a couple of turnover times. This is consistent with the fast convergence of $G_e$ for large FTLE 
(figure $\ref{F:cramerplot}$). A behavior close to this exponential increase can actually be seen in the simulations with large Prandtl numbers for 
a window of turnover times from around $4 T$ to $6 T$. Note that numerical simulations with even larger Prandtl numbers would have increased 
this time window only marginally since dividing the diffusion by two extends its time span by only half a turnover time ($\ref{E:tau_mix}$). 
The reason is that the convergence of $G_e$, at least for larger than average FTLE, has a time scale close to the advective time scale. Further 
investigations are needed to explain this fact. 

It is worth noting that the lengthening of a material contour is determined by rare events in the tail
of the FTLE distribution. The maximum $\underset{\lambda}{\max} [\lambda-G_e(\lambda,t=N T)]$ 
is achieved by values of 
$\lambda$ in the $42\%$ quantile of the distribution
for $N=2$, $27\%$ for $N=4$,  $13\%$ for $N=7$, and  $3\%$ for $N=15$.
Even though those events become exponentially rare because of the convergence of the FTLE pdfs toward a Dirac distribution, their contributions to 
the ensemble average of the contact line become exponentially important in the average of exponentials ($\ref{E:length_contact}$). 

\subsection{Lagrangian advection of the gradients along the contact line $\mathcal{L}$}

In this section, we calculate the time evolution of the gradient of $\phi$ along a Lagrangian trajectory on the contact line for infinite initial gradients.  
We take into account the time evolution of the Lyapunov exponents. The singular vectors are taken equal to the forward Lyapunov vectors:
$\psi_{+}(\mathbf{x},t)=\Psi_{+}(\mathbf{x})$.  As noted in part II.B.2, since the singular vectors converge rapidly as $t$ increases, we expect this approximation to yield
an accurate estimate of the gradients because the singular vectors 
converge rapidly, as noted in part II.B.2. 

\subsubsection{Advection-diffusion equation in a co-moving frame}

\begin{figure}
\centering
\includegraphics[scale=1.1]{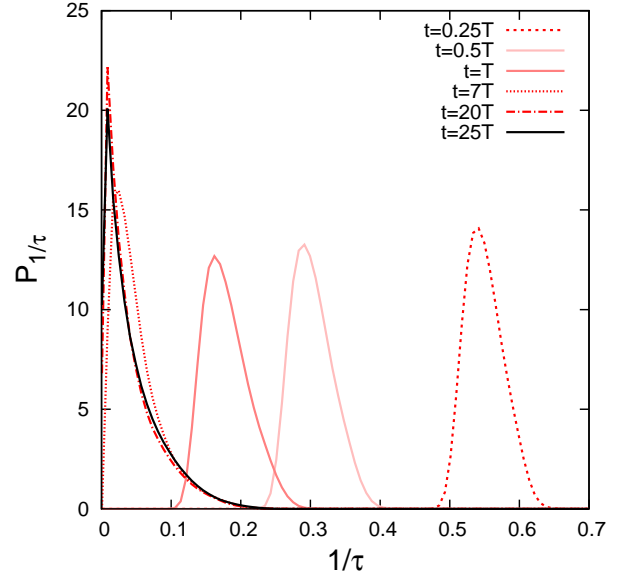}
\caption{(Color online)  \small Probability density function of $\frac{1}{\tau}$ plotted at different times. The equivalent time $\tau$ is defined in ($\ref{E:tau}$).
As the time increases, the density shifts toward smaller values.}
\label{F:plot_1tau}
\end{figure}

We consider a fluid element on the contact line and we note $\mathbf{X}_{\mathcal{L}}$ its trajectory:
\begin{align}\label{E:traj}
  \frac{d\mathbf{X}_{\mathcal{L}}}{dt}=\mathbf{u}(\mathbf{X}_{\mathcal{L}},t) & \mbox{ with } & \mathbf{X}_{\mathcal{L}}(t=0)= \mathbf{X}_{0 \mathcal{L}} \mbox{,}
\end{align}
where $\mathbf{X}_{0 \mathcal{L}}$ is the initial location of the contact line element we are following. We define a new coordinate $\mathbf{r}$ corresponding to a frame co-moving with $\mathbf{X}_{\mathcal{L}}$:
\begin{equation}\label{E:comov}
  \mathbf{r}=\mathbf{x-X_{\mathcal{L}}}\mbox{.}
\end{equation}
Writing the concentration field $\chi(\mathbf{r},t) \equiv \phi(\mathbf{x},t)$, we can show (\cite{Monin1975}) using 
($\ref{E:tracer}$) and ($\ref{E:traj}$) that:
\begin{equation}\label{E:tracer_b}
  \frac{\partial{\chi}}{\partial {t}} + [\mathbf{u}(\mathbf{X_{\mathcal{L}}+r},t)-\mathbf{u}(\mathbf{X}_{\mathcal{L}},t)]\cdot \nabla {\chi} = \kappa \nabla^2  {\chi}  \mbox{.}                                                  
\end{equation}
Assuming a separation of scale between velocity and tracer scales, we can write $[\mathbf{u}(\mathbf{X_{\mathcal{L}}+r},t)-\mathbf{u}(\mathbf{X}_{\mathcal{L}},t)]$  
at the first order in $\mathbf{r}$. We basically assume that the characteristic width of the contact zone is much smaller than the velocity scale. 
\begin{equation}\label{E:tracer_c}
  \frac{\partial{\chi}}{\partial {t}} + \mathbf{r}^T \cdot \nabla {\mathbf{u}(\mathbf{X}_{\mathcal{L}},t)} \cdot \nabla {\chi}= \kappa \nabla^2  {\chi}    \mbox{.}                                                
\end{equation}
Locally, along the contact line, the concentration of $\phi$ only varies in the direction perpendicular to the contact line, assuming that, for 
$t \lesssim T_{mix}$, the curvature of the contact line is much larger than the width of the contact zone where the gradients are concentrated.
This is relevant because the stirring in chaotic advection produces elongated structures by nature.
As a consequence, as noted previously in a similar case (\cite{Bal97}), the field $\chi$ has to be of the form: 
\begin{equation}\label{E:tracer_1D}
     \chi(\mathbf{r},t)= \widetilde{\chi}(\mathbf{k}\cdot\mathbf{r},t) =  \widetilde{\chi}(\eta,t) \mbox{,}                                         
\end{equation}
where $\mathbf{k}$ is a vector perpendicular to the contact line and $\eta$ a coordinate along $\mathbf{k}$. Substituting ($\ref{E:tracer_1D}$) into  ($\ref{E:tracer_c}$) (with
$\mathbf{S}(t) = \nabla \mathbf{u}(\mathbf{X}_{\mathcal{L}},t)$) and equaling the zero and first order terms in $\bm{r}$, we can show that (\cite{Bal97}):
\begin{subequations} 
  \begin{gather}    
   \frac{d\mathbf{k}}{dt}+ \mathbf{S}^T.\mathbf{k}=0 \label{E:wave_n}\\
   \frac{\partial{\widetilde{\chi}}}{\partial{t}}=\kappa \left|\mathbf{k} \right|^2  \frac{\partial^2{\widetilde{\chi}}}{\partial{\eta^2}} \mbox{.}\label{E:trac_rescaled}   
  \end{gather}
\end{subequations}
Equation ($\ref{E:wave_n}$) is actually the equation of a wavenumber $\mathbf{k}$ advected with the trajectory $\mathbf{X}_{\mathcal{L}}$. Noting its similarity with
($\ref{E:evol_dl}$), it is clear that the FTLE is also the maximum exponential growth rate of a wavenumber $\mathbf{k}$ 
(or equivalently of a passive tracer gradient in the absence of diffusion). This is an alternate and classical definition of FTLE (\cite{lapeyre02}). 
Considering the resolvent matrix $\mathbf{N}$ such that 
$\mathbf{k}=\mathbf{N}\mathbf{k}_0$, where $\mathbf{k}_0=k_0(-\sin \alpha, \cos \alpha)$ is the initial value of $\mathbf{k}$,
the finite time Lyapunov exponent $\lambda$ is the log of the largest eigenvalue of 
$[\mathbf{N}^{T}\mathbf{N}]^{\frac{1}{2t}}$ with $(-\sin \psi_{+},\cos \psi_{+})$ the associated eigenvector. As a consequence, we have
\begin{equation}\label{E:coordk}
\begin{array}{ll}
  \left| \bm{k} \right|^2 &= \bm{k_{0}}^{T}\mathbf{N}^{T}\mathbf{N}\bm{k_{0}}\\ &=\left | \bm{k_{0}} \right |^2 
  \left[e^{2 \lambda t}\cos^2(\psi_{+}-\alpha) + e^{-2 \lambda t}\sin^2(\psi_{+}-\alpha) \right ]\mbox{.}
\end{array}
\end{equation}
With the assumption $\psi_{+}(\mathbf{x},t)=\Psi_{+}(\mathbf{x})$,
equation ($\ref{E:trac_rescaled}$) can be written
\begin{equation}    
     \frac{\partial{\widetilde{\chi}}}{\partial{\Theta}}=\kappa   k_0^2 \frac{\partial^2{\widetilde{\chi}}}{\partial{\eta^2}}  \label{E:eq_chi}
\end{equation}
using the rescaled time
\begin{equation}
     \Theta = \left[\tau e^{2 \lambda t} \cos^2\gamma + \widetilde{\tau} \sin^2\gamma \right]  \mbox{.}\label{E:eq_time} 
\end{equation}
We have reintroduced $\gamma=\Psi_{+}-\alpha$, a random and uniformly distributed angle between $0$ and $2\pi$ (see III.A.1).
The quantities $\tau$ and $\widetilde{\tau}$ are two equivalent times defined as follows:
\begin{equation}\label{E:tau}  
    \tau = \frac{\int_0^t e^{2u\lambda(u)}du}{e^{2t\lambda(t)}} \mbox{ and } 
    \widetilde{\tau} = {\int_0^t e^{-2u\lambda(u)}du} \mbox{.}
\end{equation}
The time $\tau$, introduced by \cite{Antonsen1996} and called ``equivalent time'' by \cite{Haynes04} evaluates the stretching time scale of a 
Lagrangian parcel in the recent past because chaotic trajectories
are characterized by positive Lyapunov exponents. Similarly, the equivalent time $\widetilde{\tau}$ 
measures the stretching rate in the early part of the trajectory. As a consequence, we
expect $\tau$ and $\widetilde{\tau}$ to have the same statistics, to be asymptotically equivalent as $t \rightarrow 0$ 
and to become independent at larger times. It has been argued (\cite{Haynes04}) that the pdf of $\tau$ converges to a time asymptotic form,
which is suggested for our flow in figure $\ref{F:plot_1tau}$ where we have plotted the pdf of $\frac{1}{\tau}$ calculated
together with the Lyapunov exponent on each Lagrangian trajectory (II.2.B). The statistics of $\widetilde{\tau}$ (not shown), 
calculated the same way, are not distinguishable from these of $\tau$.

\subsubsection{Solution (infinite initial gradient case)}

The initial gradient along the contact line is infinite, while the reactants are well mixed in their respective domain with a concentration equal to
$A_0$. As a consequence, we take:
\begin{equation} \label{E:init_chi}
     \widetilde{\chi}(\eta,t=0)=A_0 \operatorname{sgn}\eta \mbox{.}
\end{equation}
The solution of ($\ref{E:eq_chi}$) with the initial condition ($\ref{E:init_chi}$) is:
\begin{equation}\label{E:sol_prof}
     \widetilde{\chi}(\eta,t)=A_0 \frac{2}{\sqrt{\pi}} \int_0^\frac{\eta}{2\sqrt{\kappa \Theta}} e^{-l^2}dl = A_0 
      \operatorname{Erf} \big(\frac{\eta}{2 k_0 \sqrt{\kappa \Theta}} \big) \mbox{.}
\end{equation}  
The function $\operatorname{Erf}$ is the Gauss error function. 
It follows from ($\ref{E:tracer_1D}$) and ($\ref{E:sol_prof}$):
\begin{eqnarray}
 \chi(\mathbf{r},t) & = & A_0\operatorname{Erf} 
			    \big(\frac{\mathbf{n} \cdot \mathbf{r}}{2 \sqrt{\kappa}} \sqrt{\frac{\left| \mathbf{k}\right|/k_0}{\Theta}} \big) \nonumber \\
		    & = & A_0\operatorname{Erf} \big(\frac{G_{\mathcal{L}}}{2 \sqrt{\kappa}} \mathbf{n} \cdot \mathbf{r}\big) \label{E:sol_prof_c}
\end{eqnarray}
with
\begin{equation}\label{E:norm}
    \mathbf{n}=\frac{\mathbf{k}}{\left|\mathbf{k}\right|}
\end{equation}  
the unit vector normal to the contact line and
\begin{eqnarray}
    G_{\mathcal{L}} 	& = & \sqrt{\frac{\left| \mathbf{k}\right|/k_0}{\Theta}} \nonumber \\
		&    =   & \sqrt{\frac{e^{2 \lambda t}\cos^2\gamma  + e^{-2 \lambda t} \sin^2\gamma }
                            {\tau e^{2 \lambda t}\cos^2\gamma +\widetilde{\tau} \sin^2\gamma }
                      } \mbox{.}\label{E:g}
\end{eqnarray}  
The norm $\left| \nabla \phi_{\mathcal{L}} \right|$ of the gradient of the field $\phi$ on the contact line (where $\chi=0$), i.e. at the location of the 
trajectory $\mathbf{X}_{\mathcal{L}}$ characterized by the Lagrangian straining properties $(\lambda,\tau,\widetilde{\tau},\gamma)$ is
\begin{equation}\label{E:grad_l}
    \left | \nabla \phi_{\mathcal{L}} \right | = \left| \nabla_{\mathbf{r}} \chi \cdot \mathbf{n} \right|_{\mathbf{r}=0}=\frac{A_0}{\sqrt{\pi \kappa}} 
						 G_{\mathcal{L}}(t,\lambda,\tau,\widetilde{\tau},\gamma) \mbox{.}
\end{equation}  

\subsubsection{Ensemble average of the gradient along the contact line}

\begin{figure*} 
\centering
\includegraphics[scale=0.9]{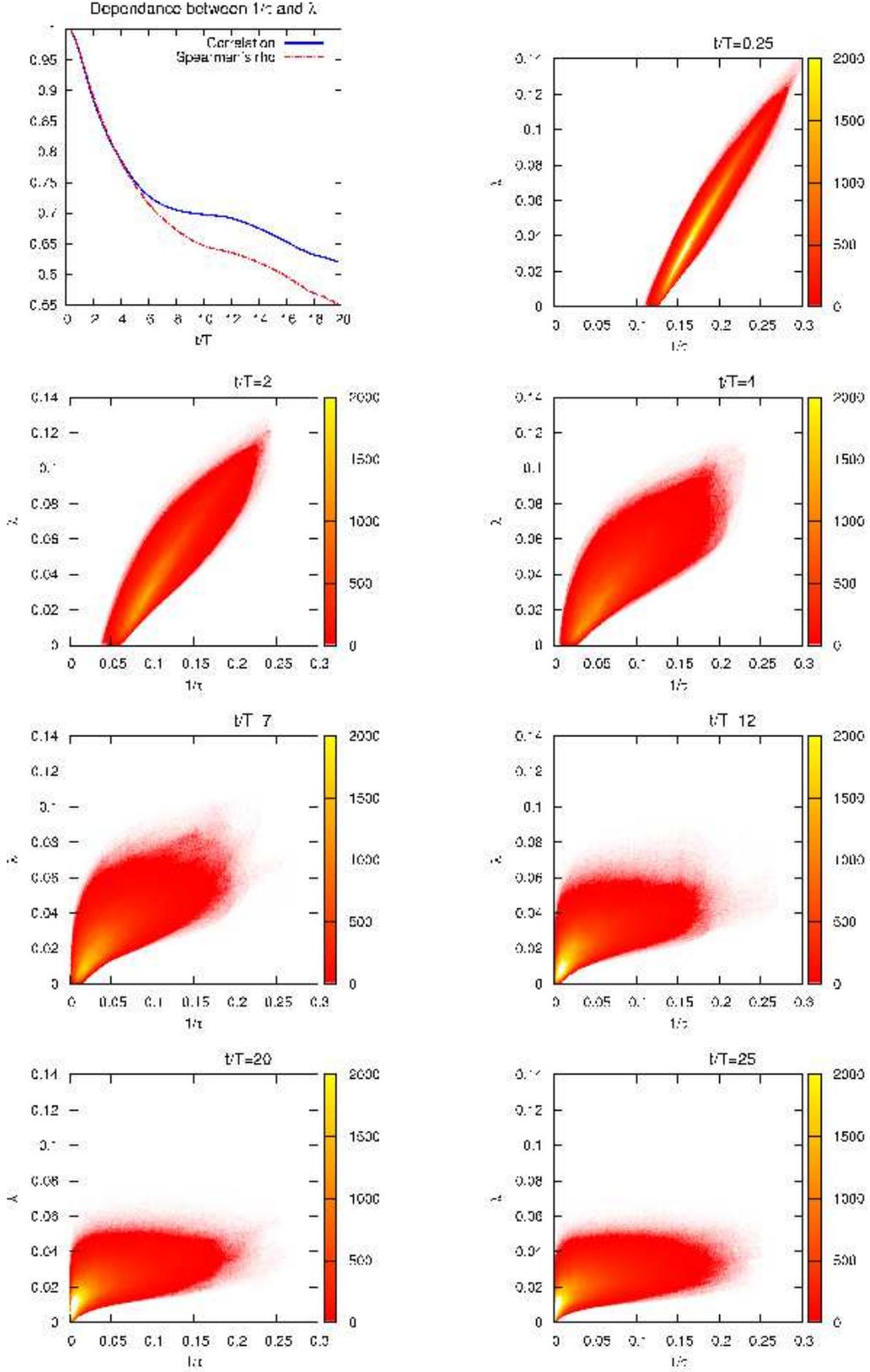}
\caption{(Color online)  \small Correlation between $\lambda$ and $\frac{1}{\tau}$ as a function of time (top left) and 
joint pdf of $(\lambda,\frac{1}{\tau})$, as estimated from the numerical simulations and plotted at different times $\frac{t}{T}=0.25,2,4,7,12,20
\mbox{ and } 25$.}
\label{F:plot_lt}
\end{figure*}

To perform the ensemble average $\langle \left| \nabla \phi_{\mathcal{L}} \right| \rangle$ of the modulus of the gradient of $\phi$ along the contact line, 
we introduce the joint pdf $\widetilde{P}$ of $(\lambda,\tau,\widetilde{\tau})$. 
As noted previously, the orientation $\gamma$ is assumed uniformly distributed between $0$ and $2\pi$ and independent from the random vector $(\lambda,\tau,\widetilde{\tau})$. 
If we consider a trajectory $\mathbf{X}_{\mathcal{L}}$ of a contact line element $\bm{\delta l}$, the gradient on it is equal to $\left | \nabla \phi_{\mathcal{L}} \right |$ 
on a length $\left| \bm{\delta l} \right|$, defined in ($\ref{E:evol_dl}$). 
As a consequence, with ($\ref{E:grad_l}$) and ($\ref{E:evol_dl}$), we obtain:
\begin{align}
\langle\left | \nabla \phi_{\mathcal{L}} \right |\rangle &= \frac{\langle \left | \nabla \phi_{\mathcal{L}} \right | \left| \bm{\delta l} \right| \rangle}
{\langle \left| \bm{\delta l} \right| \rangle} & \notag \\
&=\frac{A_0}{\sqrt{\pi \kappa}}\frac{L_0}{\langle L \rangle}
\iiiint &\frac{e^{2 \lambda t}\cos^2 \gamma +e^{-2 \lambda t} \sin^2 \gamma }
	     {\sqrt{\tau e^{2 \lambda t}\cos^2 \gamma +\widetilde{\tau} \sin^2 \gamma }} \notag \\
	&& \widetilde{P}(t,\lambda,\tau,\widetilde{\tau})\, d\lambda\, d\tau\, d\widetilde{\tau}\, \frac{d\gamma}{2\pi} \label{E:grad_ave_2}							        
\mbox{.}
\end{align}
The integration is performed between $0$ and $\infty$ for $\lambda$, $\tau$ and $\widetilde{\tau}$ and between $0$ and $2\pi$ for 
$\gamma$. Hereafter, these bounds will be omitted. For times sufficiently large ($t \gg \frac{1}{2 \langle \overline{S} \rangle} \approx T$), 
we neglect the $\sin^2$ terms under the integral in ($\ref{E:grad_ave_2}$) and in 
the expression for $\langle L \rangle$ and we obtain, in the limit of a contact line equilibrated with the flow:
\begin{equation}\label{E:grad_ave_infty_2}
\langle \left| \nabla \phi_{\mathcal{L}} \right| \rangle \underset{t \gg T} \sim
\frac{2 A_0}{\sqrt{\pi^3 \kappa}}\frac{L_0}{L_E}
    \iint \frac{e^{ \lambda t}}{\sqrt{\tau}}
	  P_{\lambda,\tau}(t,\lambda,\tau)\,d\lambda\, d\tau	\mbox{, }
\end{equation} 
where $P_{\lambda,\tau}$ is the time dependent joint pdf of $\lambda$ and $\tau$. The joint density of $(\lambda,\frac{1}{\tau})$ is pictured on figure 
$\ref{F:plot_lt}$. The frequencies $\lambda$ and $\frac{1}{\tau}$ are clearly dependent, especially when they are small, even at times
much larger than the advective time scale (e.g. $t=20T$ and $t=25T$). The computation of the Spearman Rho correlation coefficient clearly confirms this dependence 
Previous studies (e.g \cite{Antonsen1996,Haynes04}) have assumed the independence between $\lambda$ and $\tau$ at times much larger than the 
Lagrangian correlation time, here shorter than or of the order of the advective time scale. This may be apporpriate in simple ergodic chaotic flows. However, 
two dimensional Navier-Stokes flows, including two-dimensional turbulence, exhibit coherent structures (vortices, filaments of vorticity, etc...) that seem to prevent this independence 
to be achieved. 
Nevertheless, the dependence is weaker for large values of $\lambda$, which precisely dominate the integral ($\ref{E:grad_ave_infty_2}$). Approximating
$P_{\lambda,\tau}$ by the product of its marginal densities $P_{\lambda}$ and $P_{\tau}$, we obtain that 
$\left| \nabla \phi_{\mathcal{L}} \right|$ can be approximated by the simple expression $\frac{A_0}{\sqrt{\pi \kappa \tau}}$

\subsubsection{Comparison with the numerical results}

\begin{figure}[!t]
\centering
\includegraphics[scale=1.1]{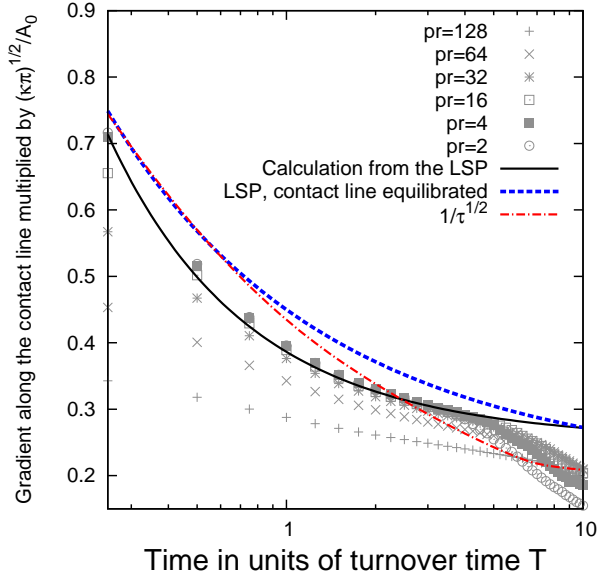}
\caption{(Color online)  \small Ensemble average of the gradients advected with the contact line, multiplied by $\frac{\sqrt{\kappa\pi}}{A_0}$, in the sharp gradient case. 
The symbols correspond to ensemble averages over the 34 DNS members 
for different Prandtl numbers $Pr=2,4,16,32,64,128$. The lines correspond to the calculation from the Lagrangian straining properties (LSP): in
solid from ($\ref{E:grad_ave_2}$) and in dashed from ($\ref{E:grad_ave_infty_2}$), considering a perfectly equilibrated contact line with the flow. The
dot-dashed line is $\frac{1}{\sqrt{\tau}}$ and corresponds to ($\ref{E:grad_ave_infty_2}$) with $\lambda$ and $\tau$ statistically independent. 
We note the log scale in the time axis.}
\label{F:plotgradf_ha}
\end{figure}

The ensemble average of the modulus of the gradient along the contact line have been calculated on the 34 ensemble members and for 
the whole range of Prandtl numbers $Pr = \frac{\kappa}{\nu}=2^i \mbox{ for } 0 \leq i \leq 7$. We calculate $\lambda$, $\tau$ and $\widetilde{\tau}$ 
on each trajectory, which permits the numerical integration of ($\ref{E:grad_ave_2}$) and ($\ref{E:grad_ave_infty_2}$). 
Numerical results are displayed on figure $\ref{F:plotgradf_ha}$ and are compared with the theoretical results of the previous paragraph. 
The joint statistics of $(\lambda,\tau,\widetilde{\tau})$ are referred to as the Lagrangian straining properties (LSP). 

The ensemble averages of the gradient calculated from the DNS and multiplied by $\frac{\sqrt{\kappa \pi}}{A_0}$ are shown for Prandtl numbers 
ranging from 2 to 128 in figure 
$\ref{F:plotgradf_ha}$. For large enough diffusion (small enough $Pr$), the curves become virtually identical, showing the dependence in 
$\kappa^{-\frac{1}{2}}$ of the gradient suggested by equation ($\ref{E:grad_ave_2}$). This regime seems to be valid for 
times up to $3.5 T$ at $Pr=2$  and up to $6 T$ at $Pr=16$. This timescale corresponds to that estimated by equation ($\ref{E:tau_mix}$) 
modulo a factor $2$ and coincides with the regime where the advection alone accounts for the lengthening of the contact line (figure $\ref{F:plotlength}$). 
The departure at small diffusion comes from the fact that the infinite gradient hypothesis becomes inaccurate in the numerical simulations given
the finite size of the grid. We can reproduce the curves at large Prandtl number by solving the derivative with respect to $\eta$ in ($\ref{E:eq_chi}$) using the 
initial condition on the gradient $\frac{\partial \chi_t}{\partial \eta}|_{_{t=0}}
=\frac{A_0}{2 \delta_0 \sqrt{\pi}} e^{-\frac{\eta^2}{4 \delta_0^2}}$, with $\delta_0$ a length corresponding to a grid point. We find that
the previous developments stand with $G_{\mathcal{L}}$  ($\ref{E:g}$) replaced by $G_{\mathcal{L},\kappa} = \sqrt{\frac{e^{2 \lambda t}\cos^2 \gamma +e^{-2 \lambda t} 
\sin^2 \gamma }{\frac{\delta_0^2}{\kappa}+[\tau e^{2 \lambda t}\cos^2 \gamma +\widetilde{\tau} \sin^2 \gamma]}}$ which
is a function of $\kappa$. The
expression $G_{\mathcal{L}}$ is a good approximation of $G_{\mathcal{L},\kappa}$ when the initial gradients imposed by the grid $\frac{A_0}{\delta_0}$ are large compared 
to $\frac{A_0}{\sqrt{\kappa \tau}}$ ($\sqrt{\kappa \tau}$ can be interpreted as the diffusive cutoff). This is not the case for $Pr=64$ and $Pr=128$ in our simulations.

The evolution of $\frac{\sqrt{\kappa \pi}}{A_0} \langle\left | \nabla \phi_{\mathcal L} \right |\rangle$ estimated using ($\ref{E:grad_ave_2}$) with the LSP is
provided on figure $\ref{F:plotgradf_ha}$. It captures very well the behavior of ensemble mean gradients for small Prandtl numbers. 
A very slight underestimation is seen that could be due to numerical artifacts or to our approximation taking the singular vectors constant in the 
theoretical developments. We also show $\frac{\sqrt{\kappa \pi}}{A_0} \langle \left | \nabla \phi_{\mathcal L} \right |\rangle$ approximated by
($\ref{E:grad_ave_infty_2}$). It overestimates the gradients at small times, a discrepancy which decreases with time as the contact line equilibrates 
with the flow. The quantity  $\frac{1}{\sqrt{\tau}}$, also shown in figure $\ref{F:plotgradf_ha}$ neither performs well at small times for the same reason as ($\ref{E:grad_ave_infty_2}$), 
nor at larger times because of the missing dependence of $\tau$ with $\lambda$.

\subsection{Time evolution of $\langle \frac{d \overline{\left | \phi \right |}}{dt} \rangle $}

\subsubsection{Theory}

Having formulated the evolution of the contact line and the gradient, we can now express the chemical speed
$-\langle \frac{d \overline{\left | \phi \right |}}{dt} \rangle $ by ensemble averaging ($\ref{E:difflux}$). We use the expression of 
$|\bm{\delta l}|$ in ($\ref{E:coorddl}$), with $\psi_{+}=\Psi_{+}$, for $dl$ and the expression $\left | \nabla \phi_{\mathcal{L}} \right |$ 
in ($\ref{E:grad_l}$) for $\left | \nabla \phi \right |$.
\begin{align}
 -\langle \frac{d \overline{\left | \phi \right |}}{dt} \rangle 
&=\frac{L_0 A_0}{\sqrt{\pi} \mathcal{A}} \sqrt{\kappa} 
\left\langle 
  \frac{e^{2 \lambda t}\cos^2\gamma + e^{-2 \lambda t}\sin^2\gamma}
  {\sqrt{\tau e^{2 \lambda t}\cos^2\gamma+\widetilde{\tau} \sin^2\gamma}} 
\right\rangle &\notag 
\\
&= \frac{L_0 A_0}{\sqrt{\pi} \mathcal{A}} \sqrt{\kappa}
\iiiint \frac{e^{2 \lambda t}\cos^2 \gamma +e^{-2 \lambda t} \sin^2 \gamma }
	    {\sqrt{\tau e^{2 \lambda t}\cos^2 \gamma +\widetilde{\tau} \sin^2 \gamma }} \notag
	   \\& \widetilde{P}(t,\lambda,\tau,\widetilde{\tau})\,
	     d\lambda\, d\tau\, d\widetilde{\tau}\, \frac{d\gamma}{2\pi}\,\label{E:tot_chim} 
\\
&\underset{t \gg T}{\large{\sim}}  \frac{2 L_0 A_0}{\sqrt{\pi^3} \mathcal{A}} \sqrt{\kappa}
\iint\frac{e^{2 \lambda t}}{\sqrt{\tau}} P_{\lambda,\tau}(t,\lambda,\tau)\,  d\lambda\, d\tau\,  \mbox{.}\label{E:tot_chim_inf}
\end{align}
The chemical speed scales like $\kappa^\frac{1}{2}$, which is a direct consequence of the scaling of the gradients like $\kappa^{-\frac{1}{2}}$, 
the contact line length being independent of the diffusion in the regime considered. Indeed, comparing equation ($\ref{E:tot_chim}$) with ($\ref{E:grad_ave_2}$)
leads to the simple relationship between the ensemble means:
\begin{equation}\label{E:chim_gen}
 -\langle \frac{d \overline{\left | \phi \right |}}{dt} \rangle = \frac{\kappa}{\mathcal{A}}\langle L \rangle 
			 \langle \left | \nabla \phi_{\mathcal L} \right |\rangle\mbox{.}
\end{equation}
This relationship was actually previously justified and used to calculate the gradients ($\ref{E:grad_ave_2}$).

\subsubsection{Numerical results}

\begin{figure}
\centering
\includegraphics[scale=1.1]{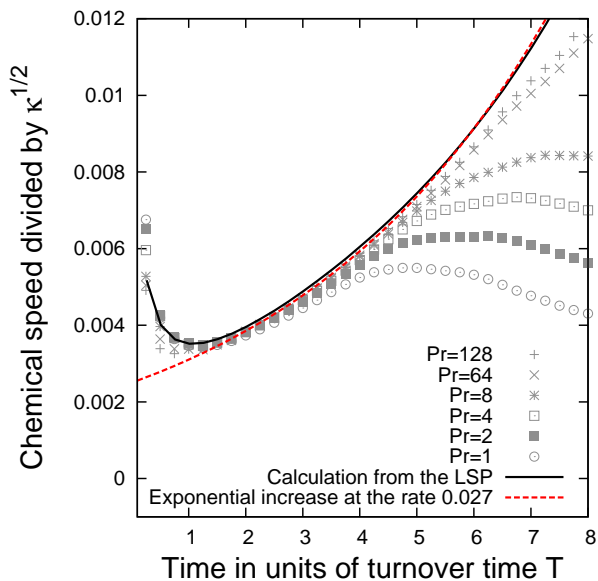}
\caption{(Color online)  \small Ensemble average of the chemical speed in the sharp gradient case divided by the diffusion $\sqrt{\kappa}$. 
The symbols correspond to numerical results from the 34 members ensemble, 
for different Prandtl numbers $Pr=1,2,4,8,64,128$. The solid line (calculation from the LSP) corresponds to ($\ref{E:tot_chim}$).
The exponential increase at a rate $\lambda_1=0.027$ (dashed line) corresponds to the expected asymptotic regime of ($\ref{E:tot_chim}$).}
\label{F:decay_ha}
\end{figure}

Figure $\ref{F:decay_ha}$ shows $-\frac{1}{\sqrt{\kappa}}\langle \frac{d \overline{\left | \phi \right |}}{dt} \rangle$ for various Prandtl numbers 
estimated from the ensemble DNS and the result of equation ($\ref{E:tot_chim}$) using LSP. Like for the gradients, the curves converge together when the diffusion gets 
larger, for times shorter than $T_{mix}$. The limit curve best fulfills the infinite gradient hypothesis and consequently matches very well the estimate 
from equation ($\ref{E:tot_chim}$) calculated from the Lagrangian straining properties. 

The general behavior of the chemical speed can be interpreted in light of equation ($\ref{E:chim_gen}$). The initial decrease is mainly due to the 
decrease of the gradients, as observed previously. Then, it is dominated by the increase of the contact line, the gradients decreasing very slowly.
Figure $\ref{F:decay_ha}$ shows the exponential increase at a rate $\lambda_1 \approx 0.027$.
The timescale corresponding to the minimum of the chemical speed can be estimated from the timescale of the decrease of the gradient, which is of the
order of $T$ \footnote[1]{Assuming that it is the time scale for the decrease of $\frac{1}{\sqrt{\tau}}$, this estimate is obtained by direct calculation of 
$\frac{1}{\sqrt{\tau}}$ from ($\ref{E:tau}$) taking $\lambda \approx S$. The latter approximation is justified because the Lyapunov exponent is very close to 
the strain rate where the trajectory originates for times smaller than $T$, as shown on figures $\ref{F:mapFTLE}$ and $\ref{F:plotpdf}$. 
The fact that we find a time scale of the order of $T$ validates the approximation $\lambda \approx S$ a posteriori.}

\subsection{Alternative initial condition on the tracers: smooth gradients}

The following calculations are extending the analytical results for an initial condition on the tracers with smooth gradients and are
validated numerically with the initial condition ($\ref{E:init-trac_2}$). We neglect the diffusion to determine the evolution of the 
gradients. In the inviscid limit, a gradient along a Lagrangian trajectory obeys the wavenumber equation ($\ref{E:wave_n}$) (e.g. \cite{lapeyre02}), 
whose solution is given by ($\ref{E:coordk}$). Together with ($\ref{E:coorddl}$), we obtain the ensemble average of ($\ref{E:difflux}$):
\begin{align}
-\langle\frac{d \overline{\left | \phi \right |}}{dt} \rangle &= 
\beta \kappa \iint  P_{\lambda}(t,\lambda)  
\left[ e^{2 \lambda t}\cos^2\gamma+e^{-2 \lambda t}\sin^2\gamma \right] \frac{d\gamma}{2\pi}\, d\lambda & \label{E:chim_cos_1} \\ 
 & \mathop{\Huge \sim}_{t \gg \frac{T}{2}}  \frac{\beta}{2} \kappa \int 
P_{\lambda}(t,\lambda) e^{2 \lambda t} d\lambda  \propto e^{\max_{\lambda} [2\lambda-G_e(\lambda,t)] t} & \label{E:chim_cos_2} \\
&\asymp e^{\lambda_2 t}\mbox{.}& \notag
\end{align}
with
\begin{equation}\label{E:lambda2}
\lambda_2=\max_{\lambda} [2 \lambda-G(\lambda)]
\end{equation}
the Legendre transform of $G$ evaluated in two and $\beta=\frac{L_0 \langle \left| \nabla \phi_{\mathcal{L}} \right| \rangle (t=0)}{\mathcal{A}}$.

\begin{figure}
\centering
\includegraphics[scale=1.1]{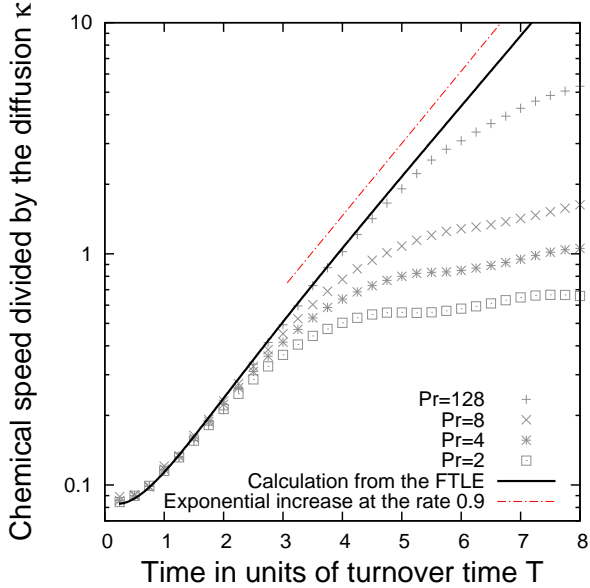}
\caption{(Color online)  \small Ensemble average of the chemical speed, in the smooth gradient case, divided by $\kappa$.
The symbols correspond to numerical results from the 34 members ensemble, 
for different Prandtl numbers $Pr=2,4,8,128$. The solid line (calculation from the FTLE) correspond to ($\ref{E:chim_cos_1}$).
The exponential increase at a rate 0.09 corresponds to the expected asymptotic regime of ($\ref{E:chim_cos_1}$), as expressed in ($\ref{E:chim_cos_2}$) and
has been shifted vertically for clarity. We note the log scale in the y-axis.} 
\label{F:prod_cos}
\end{figure}

The dependence of the chemistry on the diffusion is, like in the sharp gradient case, algebraic but the exponent 
is now 1. Our numerical simulations are consistent with this prediction: figure $\ref{F:prod_cos}$ shows
the chemical speed divided by the diffusion. For small times, all the curves are virtually identical, which confirms 
the $\kappa$ dependence of the chemical speed. 

The calculation of $-\langle\frac{d \overline{\left | \phi \right |}}{dt} \rangle$ using ($\ref{E:chim_cos_1}$) with the
pdf $P_\lambda$ reproduces very well the initial increase of the chemical speed, as shown on figure $\ref{F:prod_cos}$. Interestingly, the chemical speed has now 
a similar evolution as the contact line (figure $\ref{F:plotlength}$). Indeed, equations ($\ref{E:length_contact}$) and ($\ref{E:chim_cos_1}$)
are very similar. The quantity integrated over the density of $\lambda$ is just squared in ($\ref{E:chim_cos_1}$) compared to ($\ref{E:length_contact}$).
Using our numerical estimate of the Cramer function through ($\ref{E:lambda2}$), $\lambda_2$ can be estimated
at $0.09$, which is about three times $\lambda_1$. The chemical speed increases much faster than twice the contact line, 
which would be the case for a uniform Lyapunov exponent.
In the sharp gradient case, the chemical speed rather scales like $e^{\lambda_1 t}$, because of the action of diffusion on the gradient. This
suggests a smaller chemical speed, which may be surprising since the chemical speed, controlled by a diffusive flux, is expected to be larger when the gradients
are sharper. Actually, the chemical is not larger than in the sharp gradient case, precisely 
because of the difference in the initial gradients magnitude, but it increases much faster. 

\section{Concluding remarks}

We have studied an infinitely fast bimolecular chemical reaction in a two-dimensional Navier-Stokes flow at moderate Reynolds number with chaotic advection. 
The computation of the probability distribution function of the Lyapunov exponents 
suggests that large deviation theories may be relevant to describe its behavior after a few turnover times. 
We defined $G_e(\lambda,t)$ such that the FTLE pdf scales like $e^{-tG_e(\lambda,t)}$ and $\underset{\lambda}{\min} G_e(\lambda,t)=0$. The function
$G_e$ satisfactorily converges to a Cramer function $G$ in a couple of turnover times, at least for exponents larger than their mean value. 

\paragraph*{}
We have investigated the early regime ($\approx 5$ turnover times of the flow) of the reaction, corresponding to the time window where the contact line is 
a clearly defined material line that does
not depend on diffusion. We postulate that this time window is limited by the mix-down time scale from the large scales to the diffusive cutoff and scales like
the log of the Peclet number. 
We have related, both theoretically and numerically, the Lagrangian straining properties of the flow, as captured by the joint pdf of
the Lyapunov exponents $\lambda$ and two equivalent times $\tau$ and $\widetilde{\tau}$ ($\ref{E:tau}$), to the following quantities:

\begin{itemize}
\item \textbf{The ensemble average contact line length between the reactants $\langle L \rangle$.} After a brief transient corresponding to the equilibration 
of the contact line with the flow, i.e. to the alignment of the contact line elements with the direction corresponding to the maximum growth, independent 
of its initial orientation, the contact line lengthens like $e^{\underset{\lambda}{\max}[\lambda-G_e(\lambda,t)] t}$ which converges in time to 
$e^{\lambda_1 t}$, where $\lambda_1$ is the Legendre transform of $G$ evaluated in one and is determined by rare large events in the FTLE distribution. 

\item \textbf{The ensemble mean of the gradients along the contact line $\langle \left| \nabla \phi_{\mathcal{L}} \right| \rangle$.} It scales like 
$\kappa^{-\frac{1}{2}}$ and is determined by the pdf of ($\lambda,\tau,\widetilde{\tau}$) through ($\ref{E:grad_ave_2}$). The influence of 
$\widetilde{\tau}$ diminishes with time as the contact line is equilibrating with the flow. The dependence
between $\lambda$ and $\tau$ is crucial to accurately predict $\langle \left| \nabla \phi_{\mathcal{L}} \right| \rangle$. 
Our main assumption was the stationarity of the Lyapunov vectors, justified by their fast 
exponential convergence in time. It would be interesting to extend this work without this assumption to precise the conditions of its applicability.

\item \textbf{The ensemble mean chemical speed.} The chemical speed is defined as the modulus of the time derivative of the sum of the two reactants' 
mean domain concentrations. It scales like $\kappa^{\frac{1}{2}}$ in the limit of infinite initial gradients.
This scaling is consistent with 
\cite{Wonhas02} in the special case of a contact line of dimension one separating two on/off fields.
The ensemble average chemical speed is proportional to the product of $\langle L \rangle$ and $\langle \left| \nabla \phi_{\mathcal{L}} \right| \rangle$. 
Hence, an initial decrease of the chemical speed is related to the decrease of the gradients, while a later regime is dominated by the lengthening 
of the contact line and is consequently equivalent to $e^{\underset{\lambda}{\max}[\lambda-G_e(\lambda,t)] t}$. 
Both the contact line length and the chemical speed are determined by very rare events in the tail of the FTLE distribution. This points out the importance of
considering the distribution of the FTLE, which is not always taken into account in the literature (\cite{Sokolov1991,Karol05,Karol07}).
\end{itemize}

The case of smooth gradients exhibits some significant differences. The gradients increase instead of decreasing and are initially not affected by 
diffusion. The two main consequences are that the chemistry scales like $\kappa$ and increases exponentially in time at a rate determined by 
even rarer events in the tail of the FTLE distribution ($\ref{E:chim_cos_2}$).

The theory developed in this paper should allow to predict the evolution of the pdfs of the gradients along the contact line and of the passive tracer
$\phi$, which would be a very robust way to test it. This will be the subject of a future paper. Another paper in preparation which takes into account the
fractal structure of the contact line in order to investigate the intermediate 
regime, where the chemical production reaches a maximum, and the long term decay of the reactants.

Some interesting open questions about the Lagrangian properties of a two-dimensional Navier-Stokes flow have arisen from this study. 
What determines the initial time evolution of the FTLE pdf? What determines the shape 
of the Cramer function $G$? Is it possible to predict the asymptotic form of the pdf of $\frac{1}{\tau}$? More importantly, the dependence between
$\tau$ and $\lambda$, associated with the persistence of
significant probability of small FTLE, seems to be a major difference with simple prescribed flows used in the literature to study chaotic advection, which 
may not be, as a consequence, representative of dynamically consistent flows. Studying the joint pdf of $(\lambda,\tau)$ may happen to be useful to better
understand the mixing of both passive and active tracers in two-dimensional Navier-Stokes flows with chaotic advection, particularly when using Lagrangian straining theory approaches. 

\paragraph*{\textbf{Acknowledgments}}
\small
This research was supported by the Natural Sciences and Engineering Research Council of Canada (NSERC). We thank an 
anonymous reviewer that contributed very significantly to improve the theoretical development of part III.B.
\normalsize
\bibliography{lit_sav.bib}

\end{document}